# Development of Wind Power Generation Model with DFIG for Varying Wind Speed and Frequency Control for Wind Diesel Power Plant

Naresh Kumari [#1], A. N. Jha [*2], Nitin Malik [#3]
[#] Department of Electrical, Electronics and Communication Engineering,
The NORTHCAP University, Gurgaon, India
[1] nareshkumari@ncuindia.edu
[3] nitinmalik@ncuindia.edu
[*] Ex-Professor, Electrical Engineering, Indian Institute of Technology
New Delhi, India
[2] anjha@ee.iitd.ac.in

***Abstract*— The power generation with non-renewable energy sources has very harmful effects on the environment as well as these sources are depleting. On the other side the renewable energy sources are quite unpredictable source of power. The best trade-off is to use the combination of both kind of sources to make a hybrid system so that their individual power generation constraints can be overcome. The hybrid system taken for analysis in this work comprises of wind and diesel power generation systems. The complete modelling of the system has been done in MATLAB/SIMULINK environment. Doubly fed induction generator (DFIG) is used for power generation in wind power system. The modelling has been done considering the changing wind speed and varying load conditions. The mathematical models of DFIG and diesel power generator have been used to develop the simulink model which can be used for analysis of various performances of the system like frequency response and power sharing between different sources with load variation .The generating margin of DFIG is also simulated for the frequency support during varying load conditions .The generating margin is created by the control of active power output from DFIG. Also as the power demand rises the generating margin of DFIG keeps the balance between the power generation and load. Proportional Integral controller has been used for diesel generator plant for frequency control. The controller gains have been optimized with Particle Swarm Optimization technique. The proper selection of controller gains and wind power reserve help to achieve the enhanced frequency response of the hybrid system.**

## I. INTRODUCTION

The conventional sources of power are harmful for the environment as well as are depleting at fast rate. The solution of this problem lies in power generation from renewable energy sources. The wind energy is used mostly in the areas where high speed wind is available [1].The power production by the renewable energy sources is very much unpredictable so the complete dependency on them is not a viable alternative .The most suitable solution for the same is to use the both type of sources in integration which forms the hybrid power system. These hybrid systems are generally near to the load centres so the power can be generated at distribution level voltage. The generation capacity is also less for hybrid systems [2]. If the wind plant is operating with maximum power production then the generating margin cannot be used for the frequency control, although there can be economic benefits. For the optimum use of wind power the grid codes have been defined which support not only the conventional power generation sources but also the renewable sources of power .The wind system operation can be fixed speed or variable speed .Maximum wind power can be generated if the variable wind speed system is connected in the network for the production of energy.

This increase in energy due to variable wind speed can be 3-28 % depending on the wind conditions and design of variable speed wind turbine [3].Although the wind turbine system based on fixed speed can be connected directly in the grid still wind turbine with variable speed is used as it decreases the mechanical stresses, noise and ease of various power controls.

Further the models of wind turbine and DFIG should be as simple as possible so that the model showing all dynamics of the system when it has grid interface can be easily designed .The grid interface of wind farm is a complex task due to variable wind turbine speeds and thus various voltage sags in the system. The wind farm has more stable operation when DFIG is used in place of conventional synchronous generators and squirrel cage generator. The synchronous generator or squirrel cage generator draws heavy magnetising current from the grid after the transient state which can cause a large voltage drop [4].When DFIG is used voltage source converter is connected to DFIG rotor through slip rings and voltage can be controlled according to variable wind speed to give optimum power generation [5].

In the present work the frequency response has been enhanced for the hybrid power system using the reserved power of wind power plant along with the fully optimized PI controller for diesel plant. The power shared among the sources is also optimized in such a way that the generation from diesel plant is minimum and maximum power production is through wind plant. The power sharing between the two plants has been shown at two different load conditions.

## II. Wind Turbine and DFIG in Grid and Hybrid System

The general connection diagram of DFIG in grid connected mode is as shown in Fig 1[6]. The stator of DFIG is connected to three phase supply through the grid and rotor is connected through voltage source inverters through common DC link.

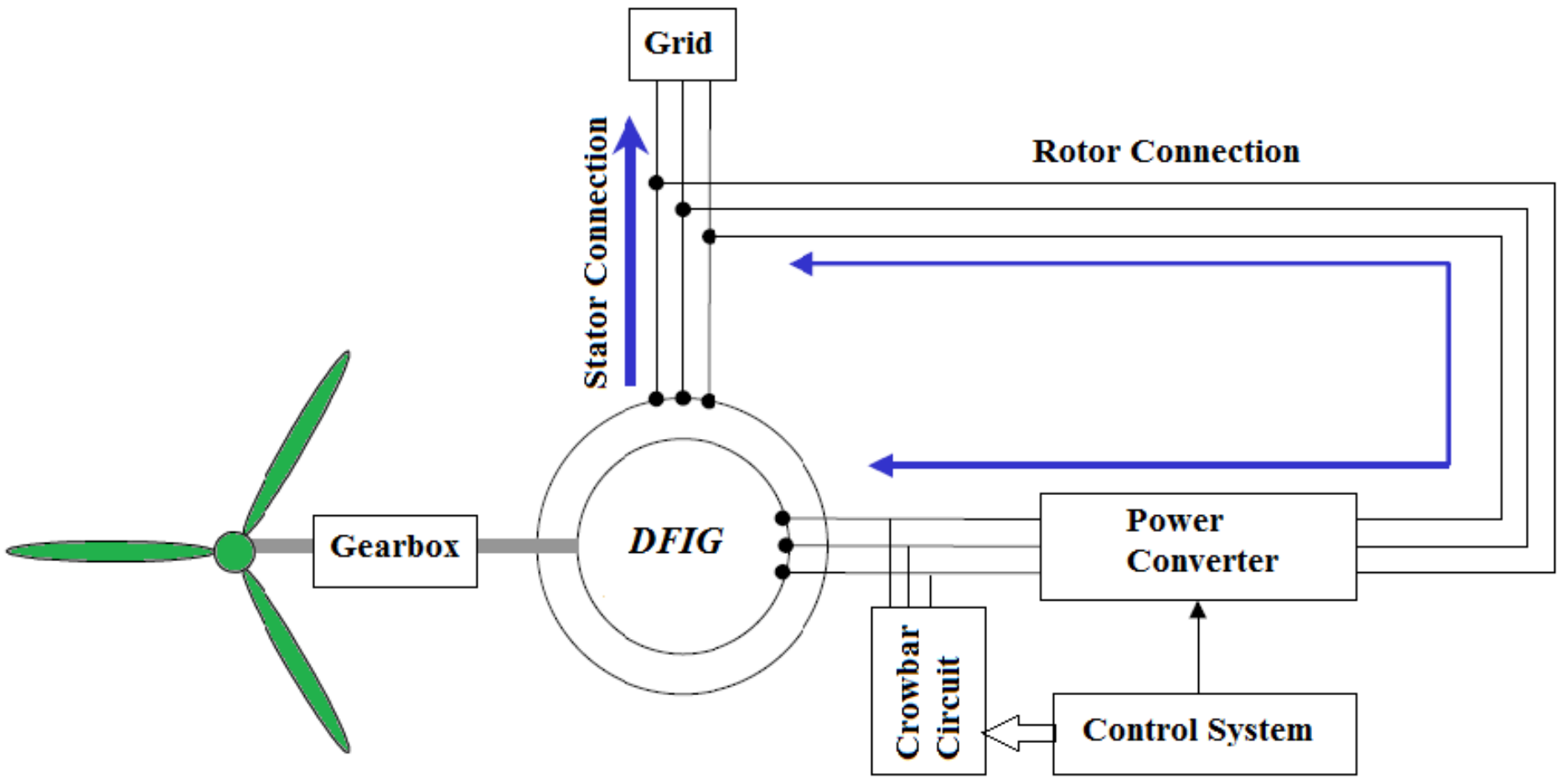

Fig. 1 General connection diagram of DFIG with Grid

When DFIG is connected to small non-renewable energy system to make hybrid system the general connection diagram can be shown as below:

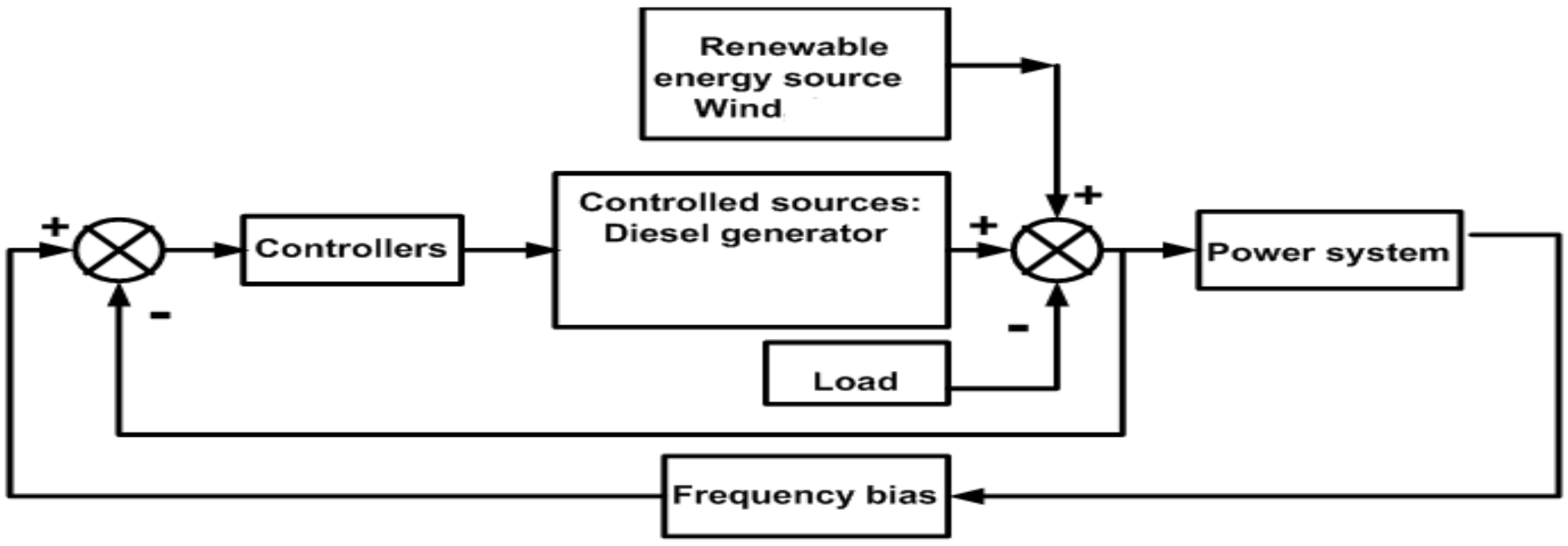

Fig. 2 General connection diagram of Hybrid Power System

In this system the PI controller can be used for controlled sources like diesel generator. The hybrid system shares the power in such a way that the load on non-renewable energy sources is minimum. Further, the DFIG also support in frequency control in such small grids by using the power from generating margin.

## III. Modelling Hybrid Power System

### *A. Modelling of Wind Power System*

When load on wind power system changes or the wind speed varies then the generating margin of the unit can be used for frequency response enhancement [7]. In case of generating margin the full power of wind plant is not used generally 10% is kept as a reserve which is used during the increase in load or reduction in wind speed. The scheme of generating margin in wind power generation is shown in Fig. 3 [8].The wind model has been designed for free running mode for medium wind speed condition when pitch angle is kept fixed. First a set point command (Pcmd) is sent to the rotor of DFIG which is added to the speed droop (ΔP) and corresponding reference signal (Pref ) is generated .This signal generates a reference torque (Tcmd ) and the inertial response adjusts the Tcmd and the corresponding signal generated will drive the DFIG which produces the Pgrid.

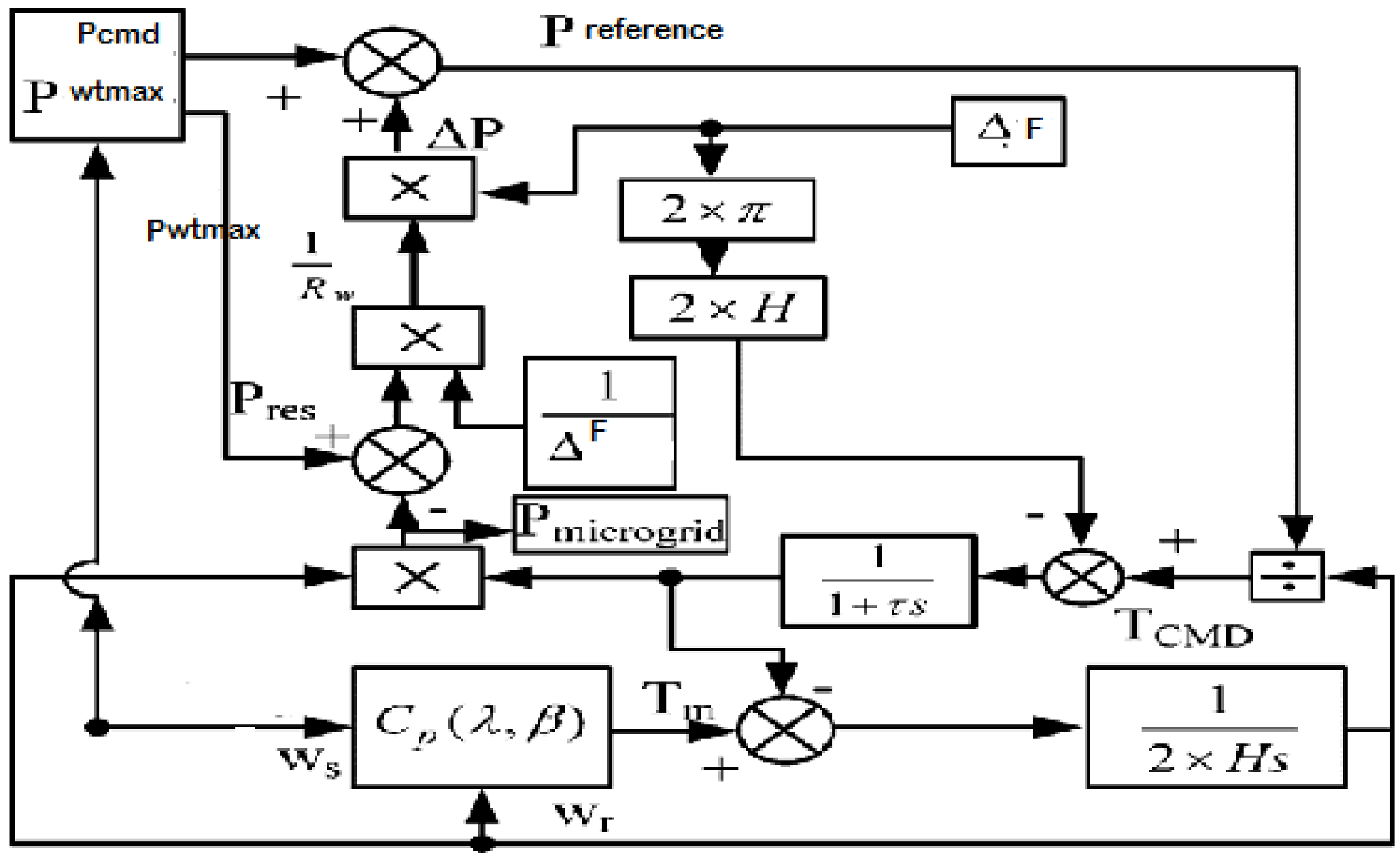

Fig. 3 Generating Margin in Wind Power Plant

The equation for maximum wind power (P wtmax) is given below:

P wtmax = 0.5 Cp(λ,β) ρAVw3 (1)

Where

- Vw Speed of wind,
- ρ Density of air,
- A Cross sectional area of the rotor of the DFIG
- Cp Power coefficient
- λ Tip speed ratio
- B Pitch angle

Cp(λ,β) of a variable speed wind turbine is :

Cp( λ , β ) =Σ Σ∝ β i λ j (2)

Where β is pitch angle and λ is tip speed ratio of wind turbine .The range of i and j is from 0 to 4 as Cp(λ,β) is a fourth order equation.

The reserve power or generating margin (Preserve) of DFIG is given as:

Preserve = Pwtmax – Pgrid (3)

The speed droop (1/Rwt) for fixed frequency band (ΔFBD) is given as:

1/Rwt= Preserve/ΔFBD (4)

The power availability (ΔP) due to speed droop depends on the variation in frequency (ΔF):

ΔP = - ΔF × 1/Rwt (5)

*B. Modelling of Diesel Generating plant*

The diesel generator plant has governor and turbine, the transfer function of these parts can be given as:

Gdg( s) = Kdg / (1+sTdg ) (6)

Gdt( s) = Kdt / (1+sTdt) (7)

- Tdg Time constant of governor
- Tdt Time constant of turbine
- Kdg Gain of governor
- Kdt Gain of turbine

The various parameters of wind generation system and diesel generator are as given in Table 1.

## IV. Simulation of Wind – Diesel generating system

The Wind –Diesel hybrid generating system has been simulated in MATLAB/SIMULINK .The part of the simulink model related to DFIG and generating margin as shown Fig. 4 is developed with the help of equations 1 to 5 as mentioned in section III .The wind speed has been considered in medium range which is generally 7.5 to 8.5 m/s. The equations 6 and 7 have been used to develop the diesel generator model in Fig 4.

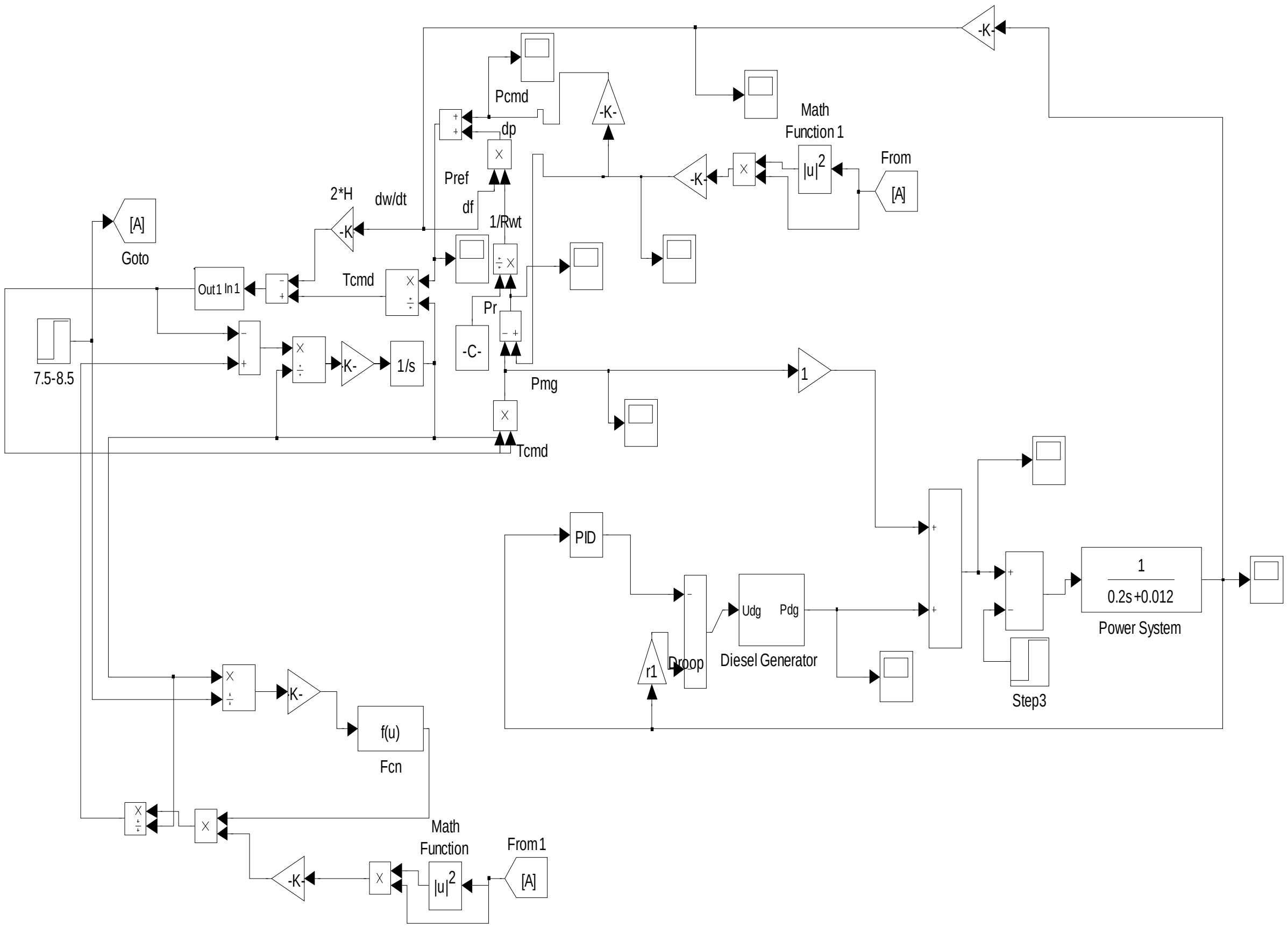

Fig 4. MATLAB/SIMULINK model for Wind – Diesel system with generating margin in wind plant

The total load is shared among the wind generation system and diesel generator system in such a way that the minimum load is on the diesel system .The generating margin is preserved in wind system and PI controller is used in diesel plant for frequency control. The active power output of the wind power plant can be changed with the d-axis rotor current control as it affects the electromagnetic torque ([9] - [11]). In the present work the pitch angle control ([12], [13]) has not been applied as the output of turbine is always more than the power requirement.

## V. Simulation Results and Discussion

The wind system has been designed in such a way that the active power output from the plant does not attain its maximum generating capacity, it helps to preserve the generating margin. The generating margin from wind power plant is possible when active power output of the wind turbine can be controlled. The generating margin helps in frequency response when the wind speed is high and wind power produced is more than the power demand. For preserving the generating margin a suitable power command to the wind turbine is given which is generally 90% of the total maximum power generation possible. The remaining 10% of power is kept as generating margin which is extracted from the system when there is some load variation or change in wind speed ([14]-[16]). The generating margin is equivalent to spinning reserve of conventional power plants. T3he simulation work has been done to observe the effect of load variation on the power generation from different units in the wind –diesel system. During the load varying conditions, the wind speed has been taken as constant.

The total generating capacity of the wind–diesel system is 350 KW. The maximum generating capacity of wind plant is 310 KW and that of diesel plant is 40 KW. The system developed is put on a total load of 295 KW and 306 KW at different time intervals as shown in Fig. 5. The power is shared among the wind and diesel power plants under the different load conditions. The wind plant generation is 275 KW and 276 KW respectively during different loads as given in Fig. 6 and diesel power generation is 20 KW and 30 KW as shown in Fig. 7. The results show that the frequency deviation is minimum during both the load variations as shown in Fig. 8. The integral square error (ISE) of the system has been taken as performance index and it becomes minimum as the frequency deviation approaches to zero as shown in Fig. 9.

In addition to frequency regulation, two more advantages are attained with this set up, first is the load shared by the non-renewable energy source i.e. diesel generator is not maximum and power generation capacity of wind plant is also not utilised to the maximum limit hence the generating margin of the wind power plant is preserved. The dependency on the fossil fuel is reduced .The system is not using any storage device and is a self-sustained system. The results show that the active power generation of wind and diesel plant is optimized during the load variation on the system and frequency can be maintained within suitable limits.

The fully optimized PI controller has been used for diesel power plant to enhance the frequency response during load variation. The proportional gain (Kp), integral gain (Ki) and governor droop (R) have been optimized with Particle Swarm Optimization (PSO) technique. The objective function (ISE) of the system taken for optimization is as given in equation (8).

$$ISE = \int \{(\Delta fi)2 + (\Delta Ptiei\text{-}j)2\}dt \quad (8)$$

The population size in PSO is taken as 100 and maximum number of iterations are taken as 150.The values of Kp, Ki and R after the optimization are 2.231, 0.0651 and 0.2273 respectively for the load of 295 KW and Kp, Ki and R for a load of 306 KW are 2.0539, 0.0655 and 0.4219 respectively.

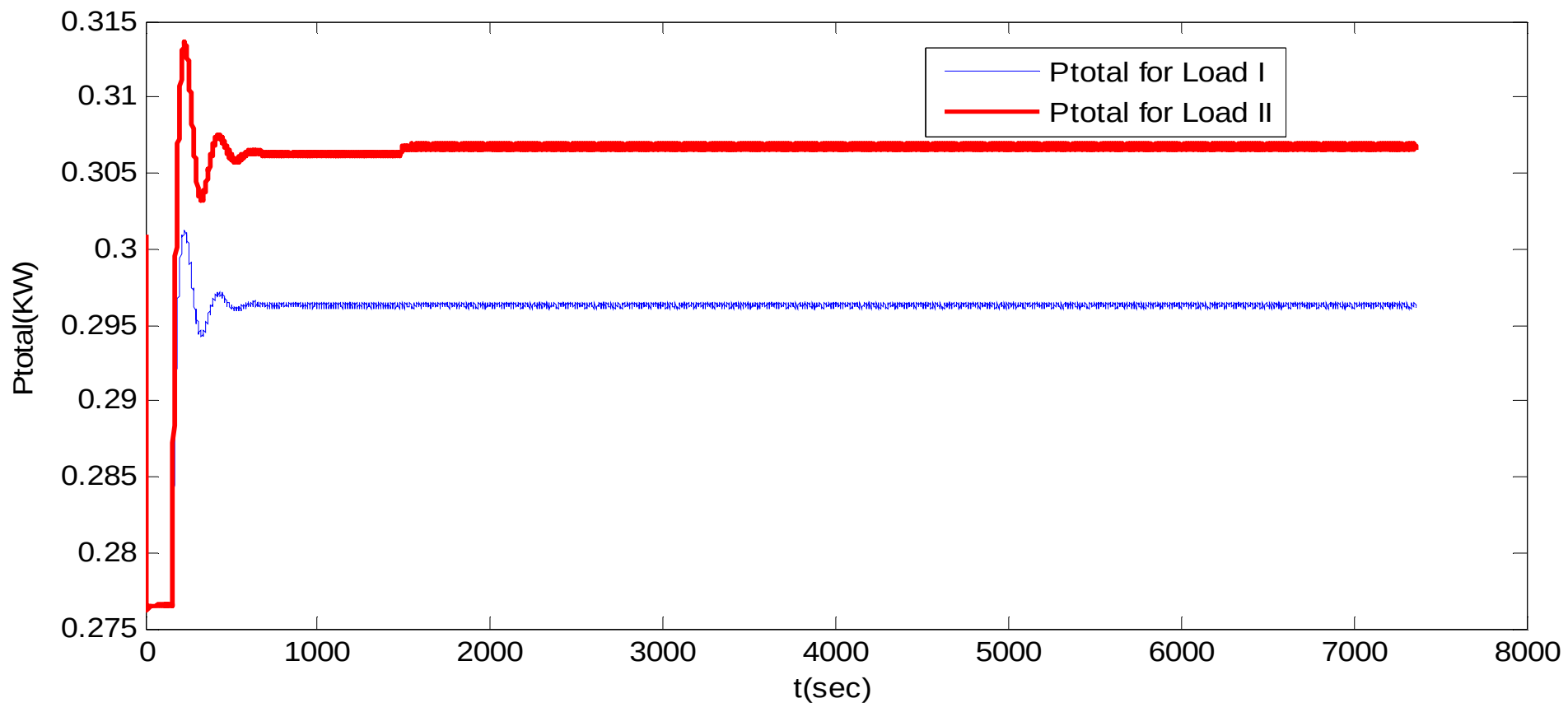

Fig. 5 Total power generated by wind and diesel power systems

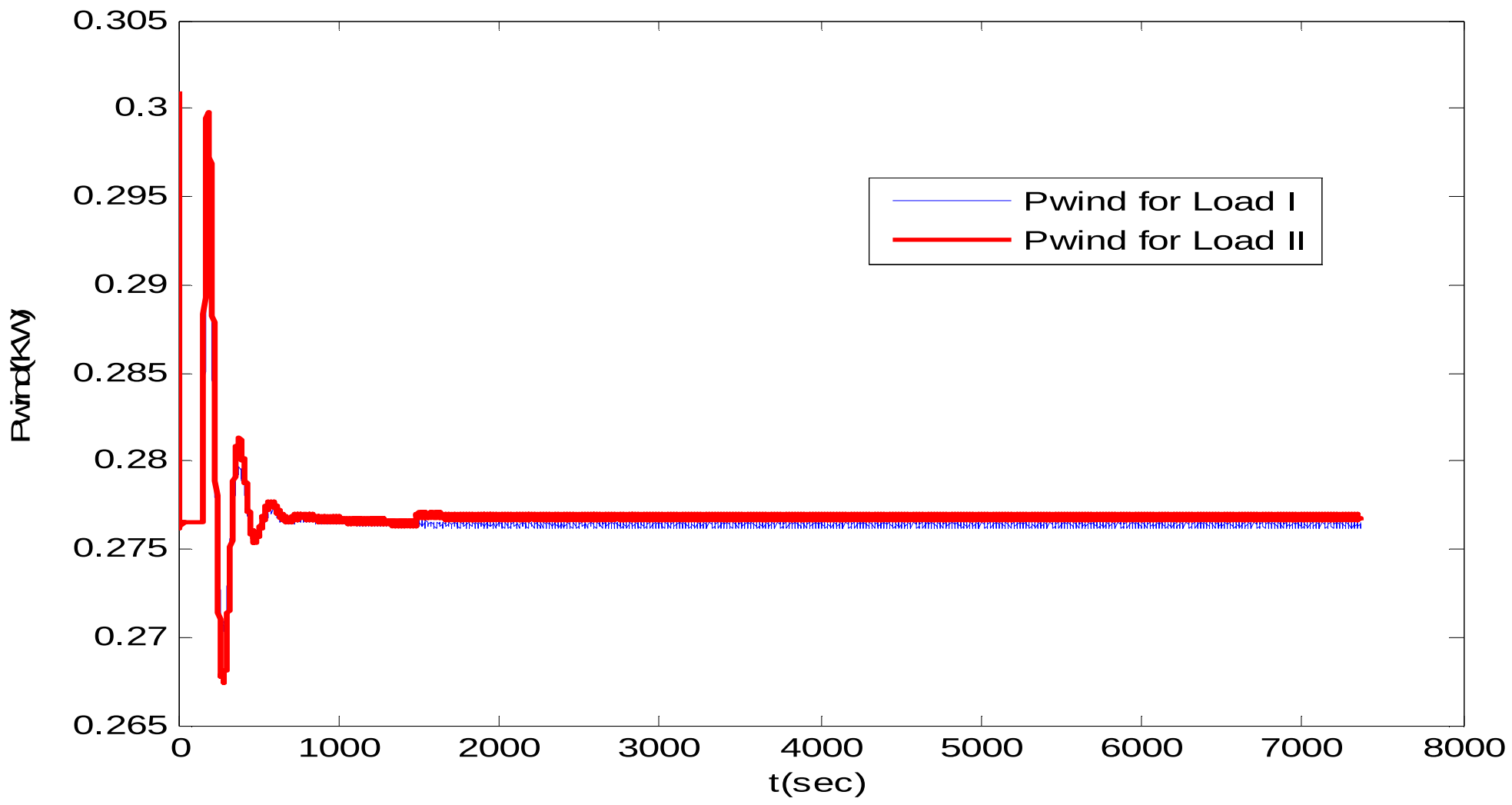

Fig. 6 Power generated by wind power plant

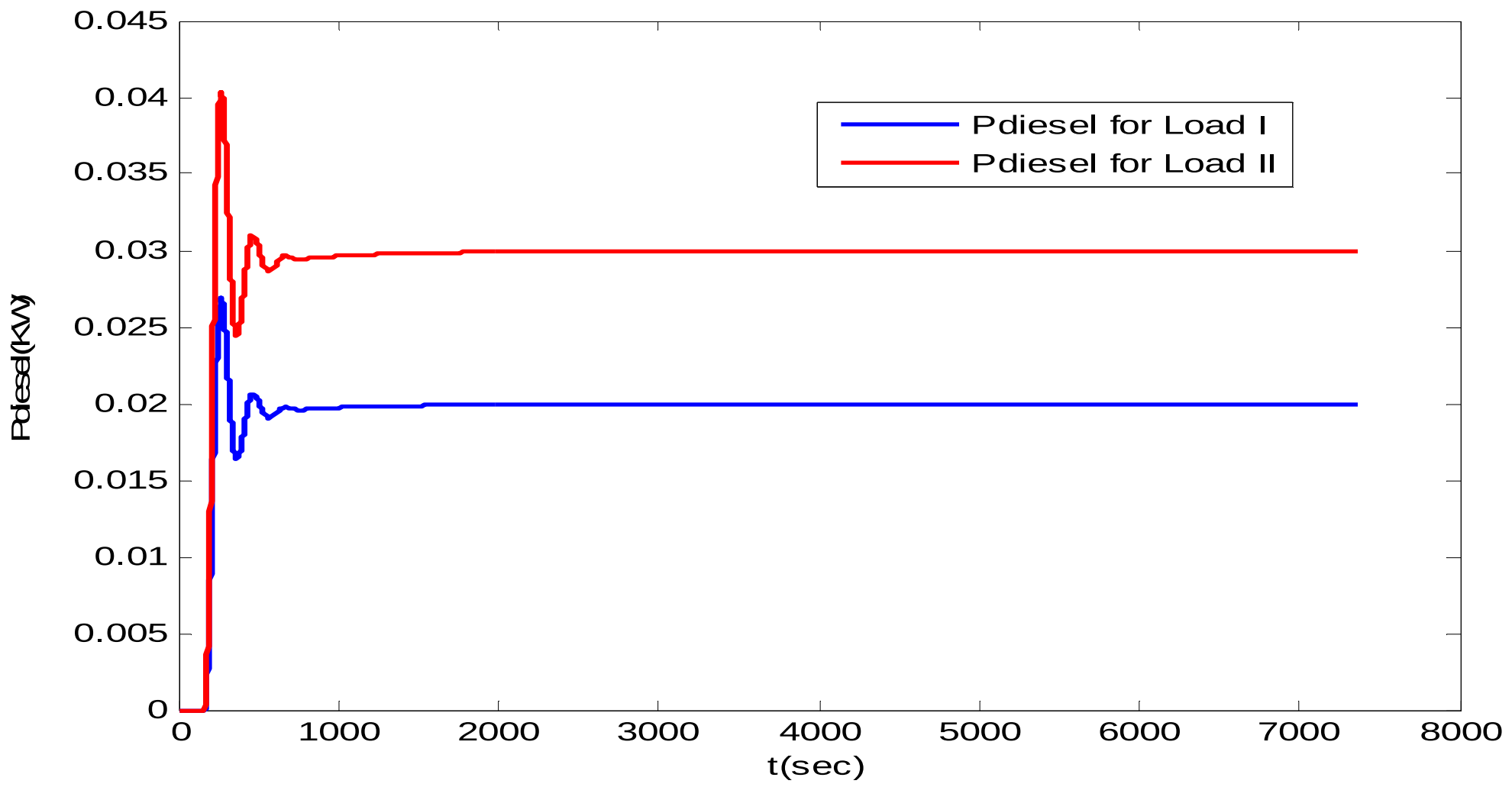

Fig. 7 Power generated by diesel power plant

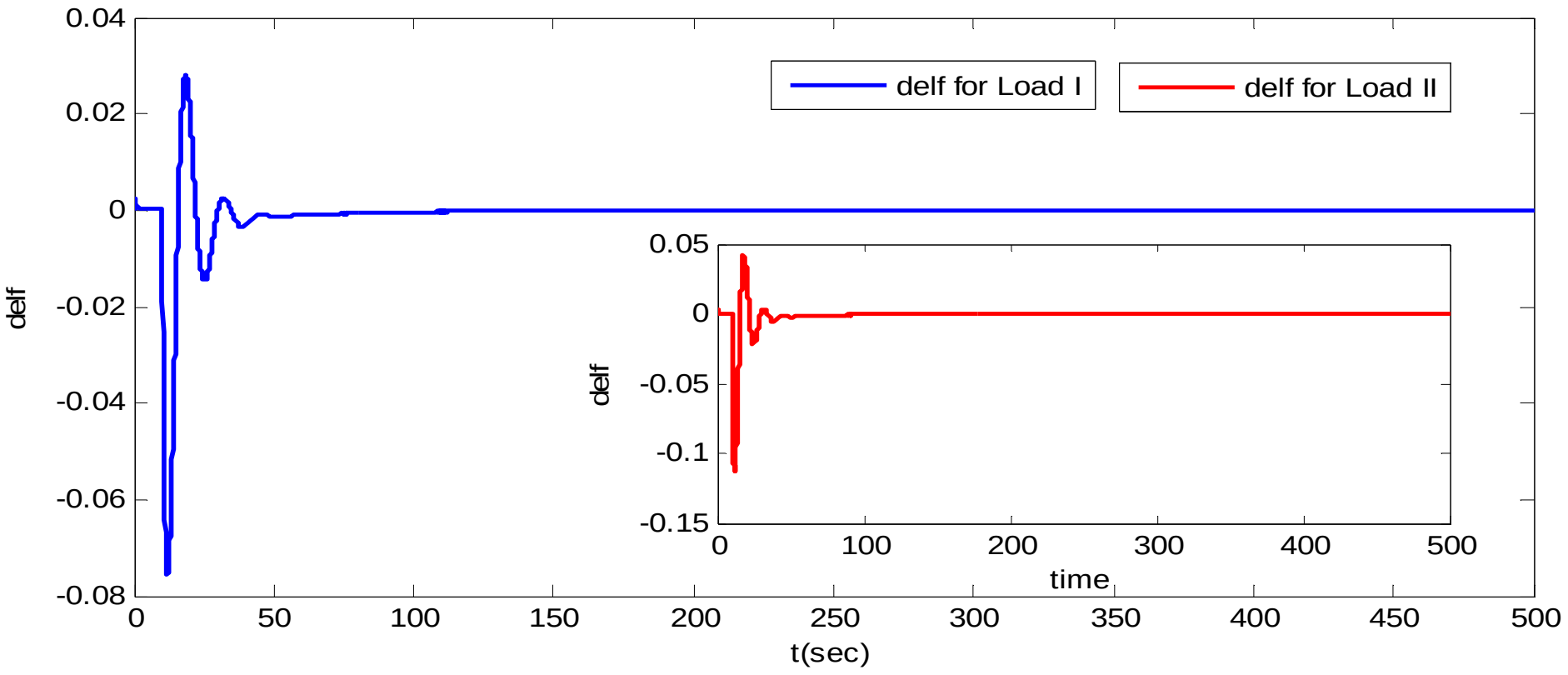

Fig. 8 Frequency variation for wind – diesel system under load variation

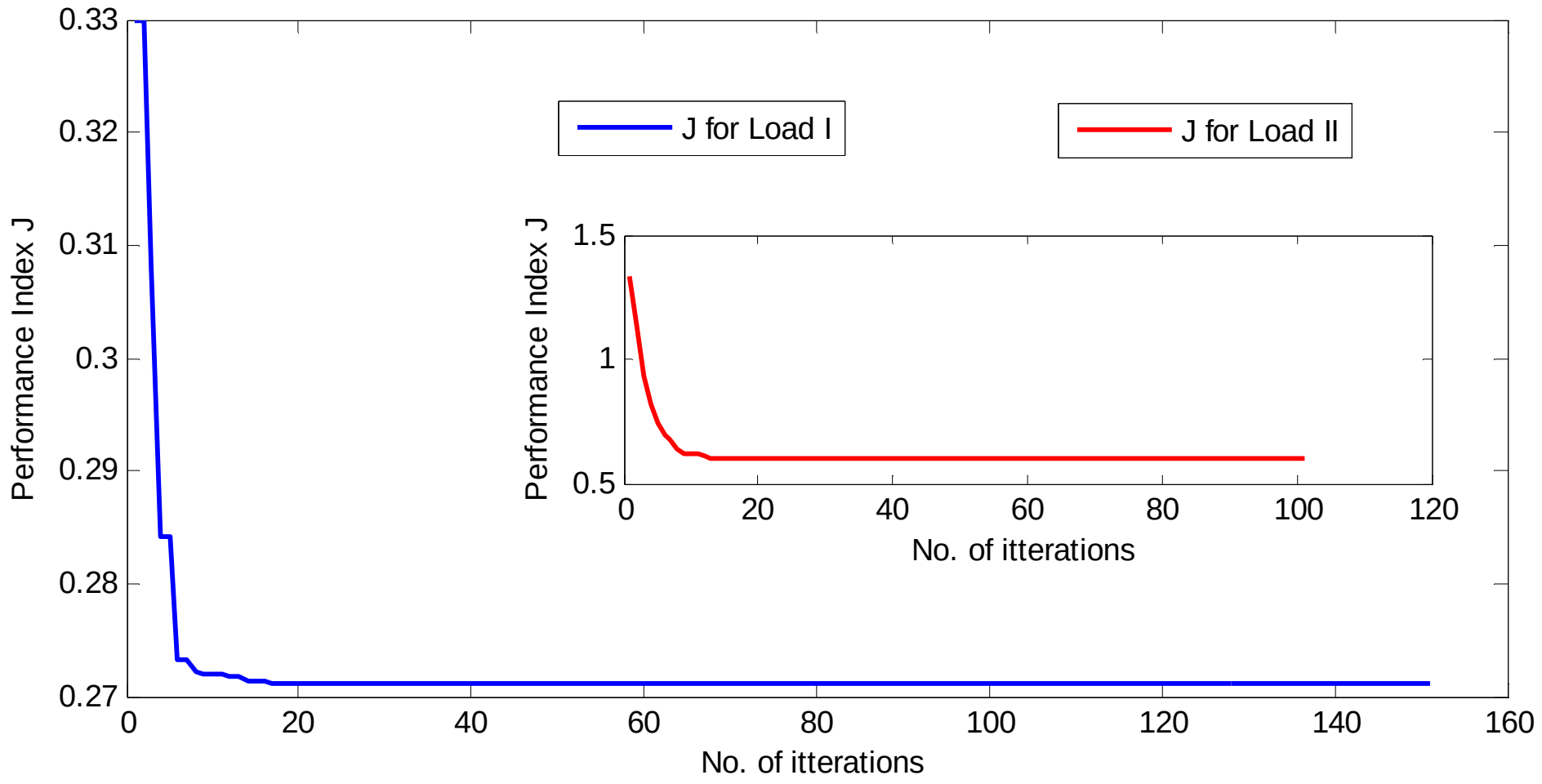

Fig. 9 Performance Index of the system under different load conditions

## VI. Conclusion

In this work the model for a doubly fed induction generator (DFIG) along with variable speed wind turbine has been developed. The wind energy conversion system is integrated with diesel generator plant which together makes a hybrid system. The combined system is connected to a common load and acts in an isolated mode .This system is designed for the remote coastal areas which cannot be connected to the main grid but they have abundance of wind energy. The problem arises when there is reduction in wind speed or increase in load on the generation system. There can be large frequency deviation if the power generation does not match with the load demand. Two methods have been implemented for frequency response enhancement of the system. Firstly the maximum power generation from wind plant is avoided and power is reserved to meet the increase in load demand Secondly the power is generated by the diesel power plant which is controlled through fully optimized PI controller and shares the load requirement with wind power generation system. This hybrid system works very effectively by optimum dependency on non-renewable energy source like diesel power plant and renewable energy source like wind power plant.

## APPENDIX

TABLE 1. Parameters of Hybrid Power System Simulated

| | |
|---|---|
| Density of air = 1.25 kg/m3<br>Gear ratio =70 | T dg = 2 s |
| Radius of turbine blade= 45 m | D = 0.012 MW/Hz |
| wind velocity= 7.5-8.5 m/s | T i (wind) = 3s |
| H=5 sec | Tpt(wind) =10 s |
| F= 50 Hz | GRC dg = 3% |
| Kdg = Kdt = 1 s | Tdt = 20 s |

## AUTHOR PROFILE

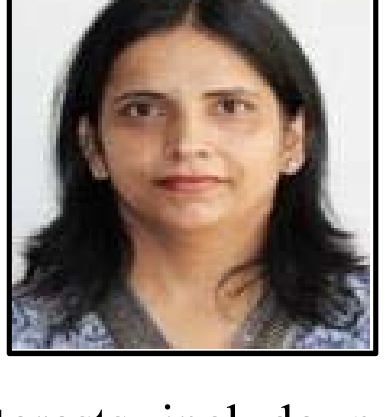

Naresh Kumari received her B.E. in Electrical Engineering with Honours in 1996 from C.R. State College of Engineering, Murthal, Haryana. She did her M. Tech. in Instrumentation and Control from Rajasthan University, Udaipur in 2005 .She is currently pursuing her PhD under the supervision of Prof. A.N.Jha and Dr. Nitin Malik from The NORTHCAP University (NCU), Gurgaon, India. She is also working as Senior Assistant Professor in Department of Electrical, Electronics and Communication Engineering, The NCU, Gurgaon. She has 18 years of teaching and research experience. Her research interests include power system stability, renewable energy sources, intelligent control and wind energy conversion systems.

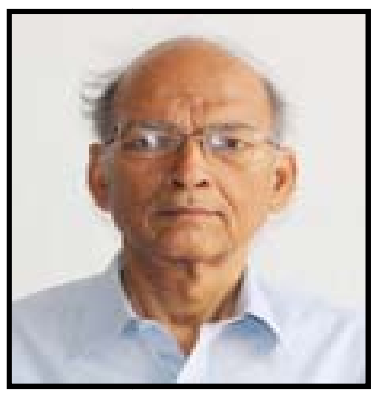

A.N.Jha has worked as Senior Professor in Department of Electrical, Electronics and Communication Engineering, The NCU, Gurgaon, India. He was Professor of Electrical Engineering in IIT, Delhi, India before joining The NCU, Gurgaon, India. He received his B.Sc. in Electrical Engineering from Bihar University, India in 1965 and M.E. (Electrical Engineering) in 1967 from University of Calcutta, India. He received his Ph.D. in Control System Engineering (Electrical) in 1977 from IIT Delhi, India. He has more than 40 years of teaching and research experience and has published/presented more than 120 papers in international and national journals/conferences. He has guided 70 M. Tech. and 7 Ph.D. students in their theses/projects.

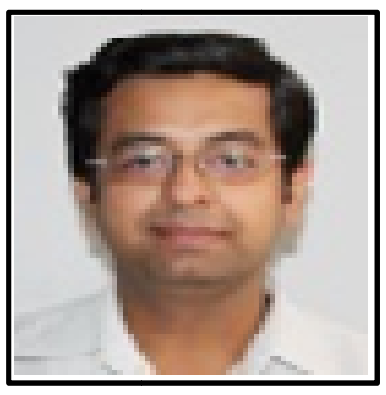

Nitin Malik received his B.Eng. (Electrical Engineering.) from CRSCE, Murthal in 1999, M.Eng. (Industrial System & Drives) from MITS, Gwalior in 2001 and Ph.D (Electrical Engineering) from Jamia Millia Islamia, New Delhi, India. He is Associate Professor at The NCU, Gurgaon, India. His research interest includes soft computing applications to Electric Power Distribution System security analysis, optimization and control.